\documentclass[aps,prb,preprint]{revtex4-1}
\usepackage{graphicx}
\usepackage{epstopdf}
\usepackage{tabularx}
\usepackage{graphicx}

\usepackage{bm}
\usepackage{color, soul}
\usepackage{multirow}
\newcommand{\Mz}{$\mathcal{M}_z$}
\newcommand{\Gz}{$\mathcal{G}_z$}
\newcommand{\etal}{{\em et al}.\ }

\begin{document}
\title{Designing Xenes with Two-Dimensional Triangular Lattice}
\author{Xu Duan}
\affiliation{Fudan University, Shanghai 200433, China}
\affiliation{School of Science, Westlake University, Hangzhou, Zhejiang 310024, China}
\affiliation{Institute of Natural Sciences, Westlake Institute for Advanced Study,Hangzhou, Zhejiang 310024, China}
\author{Zhao Liu}
\affiliation{Hefei National Laboratory for Physical Sciences at the Microscale, University of Science and Technology of China, Hefei, Anhui 230026, China}
\affiliation{School of Science, Westlake University, Hangzhou, Zhejiang 310024, China}
\affiliation{Institute of Natural Sciences, Westlake Institute for Advanced Study,Hangzhou, Zhejiang 310024, China}
\author{Brendan M Hanrahan}
\affiliation{Sensors \& Electron Devices Directorate, U.S. Army Research Laboratory, Adelphi, Maryland 20783, United States.}
\author{Wei Zhu}
\affiliation{School of Science, Westlake University, Hangzhou, Zhejiang 310024, China}
\affiliation{Institute of Natural Sciences, Westlake Institute for Advanced Study,Hangzhou, Zhejiang 310024, China}
\author{Shi Liu}
\email{liushi@westlake.edu.cn}
\affiliation{School of Science, Westlake University, Hangzhou, Zhejiang 310024, China}
\affiliation{Institute of Natural Sciences, Westlake Institute for Advanced Study,Hangzhou, Zhejiang 310024, China}


\begin{abstract}{
Xenes, graphene-like two-dimensional (2D) monoelemental crystals with a honeycomb symmetry, have been the focus of numerous experimental and theoretical studies. In comparison, single-element 2D materials with a triangular lattice symmetry have not received due attention. Here, taking Pb as an example, we investigate the triangular-lattice monolayer made of group-IV atoms employing first-principles density functional theory calculations. The flat Pb monolayer supports a mirror-symmetry-protected spinless nodal line in the absence spin-orbit coupling (SOC). The introduction of an out-of-plane buckling creates a glide mirror, protecting an anisotropic Dirac nodal loop. Both flat and buckled Pb monolayers become topologically trivial after including SOC. A large buckling will make the Pb sheet a 2D semiconductor with symmetry-protected Dirac points below the Fermi level. The electronic structures of other group-IV triangular lattices such as Ge and Sn demonstrate strong similarity to Pb. We further design a quasi-3D crystal PbHfO$_2$ by alternately stacking Pb and 1T-HfO$_2$ monolayers. The new compound PbHfO$_2$ is dynamically stable and retains the properties of Pb monolayer. By applying epitaxial strains to PbHfO$_2$, it is possible to drive an insulator-to-metal transition coupled with an anti-ferroelectric-to-paraelectric phase transition. Our results suggest the potential of  the 2D triangular lattice as a complimentary platform to design new type of broadly-defined Xenes.}
\end{abstract}
\maketitle
\newpage

\section{Introductions}
Since the discovery of  monolayer graphene and its demonstrated superb electrical, thermal, and mechanical properties, two-dimensional (2D) materials have attracted tremendous attention, driven by their potential applications for high-performance electronics~\cite{Houssa16p17,Balleste11p20}. The atomic thickness of 2D materials is advantageous for device scaling as well as energy efficiency by lowering the operating voltage~\cite{Chhowalla16p16052}. Distinct from 3D (bulk) materials, the surface of 2D materials does not suffer from dangling bonds, rendering them great flexibility to build up van der Waals (vdW) heterostructures in a highly controlled fashion less impacted by lattice mismatching. Moreover, vdW heterostructures can exhibit emergent phenomena that are 
absent in the constituent 2D materials, offering a rich platform to create and manipulate novel physical and chemical properties~\cite{Brihuega12p196802,Geim13p419,Novoselov16p6269,Liu16p9,Cao18p43}.  

The family of 2D materials has expanded considerably since the exfoliation of graphene via the Scotch tape technique~\cite{Novoselov04p5696}, now including materials  systems possessing diverse electrical properties ranging from metals and semiconductors to wide-bandgap insulators~\cite{Novoselov05p10451,Randviir14p426}. Such remarkable versatility opens up the opportunity for the design and development of flexible, ultrathin electronics. Notable 2D systems are graphene, black phosphorus, transition-metal dichalcogenides such as MoS$_2$, and hexagonal-BN (h-BN), which are the subjects of extensive investigations~\cite{Bridgman14p1344,Radisavlievic11p147,Liu03p104102,Memon14p5060,Saha15p14211,Wu18p1700491,Joensen86p457}.  First-principles-based high-throughput calculations have identified thousands of exfoliable compounds, offering useful guidelines for experimental synthesis of new 2D systems~\cite{Choudhary17p5179,Mounet18p246}. Roy \etal recently built a field-effect transistor with only 2D materials, using MoS$_2$ as the channel material, h-BN as the top-gate dielectric, and graphene as the source/drain and the contacts~\cite{Roy14p6259}. 

The graphene-like single-element 2D materials referred to as Xenes hold an important place in the 2D family~\cite{Grazianetti20p1900439}. After the discovery of graphene, it was only natural to explore the possibility of fabricating honeycomb-lattice monolayers consisted of other group-IV elements. Given the dominant role of silicon in the electronics industry, one major technology driver for the Xene development is to enable field-effect transistors based on silicene,  the silicon-based analogue of graphene~\cite{Patrick12p155501,Kara12p1,Tao15p227,Salahuddin18p442}. More recently, Kane and Mele pointed out that the intrinsic spin-orbit coupling (SOC) of carbon atoms, albeit weak, opens a tiny gap in the band structure of graphene, converting the system from a Dirac semimetal to a 2D topological insulator (TI) with quantum spin hall (QSH) effect characterized by counter-propagating edge currents with opposite spins~\cite{Kane05p226801}. However, the small band gap ($\approx$10$^{-3}$ meV) in graphene~\cite{Yao07p041401} makes it difficult to take advantage of the helical conducting edge states for spintronic applications.  Xenes made of heavy group-IV atoms with strong SOC effects are therefore promising candidates for room-temperature QSH insulators~\cite{Ezawa15p121003}. To date, all group-IV Xenes (silicene, germanene, stanene, and plumbene) have been synthesized experimentally~\cite{Patrick12p155501,Lin12p045802, Dvila14p095002, Acun15p443002,Zhu15p1020,Yuhara19p1901017,Balendhran15p640}. Different from graphene which is ideally flat, Xenes based on Si, Ge, Sn or Pb atoms exhibit buckled hexagonal honeycomb structures in their most stable forms. Interestingly, only plumbene is predicted to be a trivial insulator instead of a 2D TI, though the nontrivial topological states can emerge through electron doping~\cite{Zhao16p1,Huang14p105018,Yu17p125113}. It is noted that the concept of Xenes has become more general~\cite{Grazianetti20p1900439}, now also refering to monoelemental 2D sheets comprised of elements around group IV such as borophene~\cite{Mannix15p1513, Feng16p563}, phosphorene~\cite{Zhang16p4903}, bismuthene~\cite{Reis17p287}, gallenene~\cite{Kochat18pe1701373}, and arsenene~\cite{BeladiMousavi18p1807004}. 

Given the supreme status of graphene in 2D material research, it is not surprising that most studies on single-element 2D systems focused on the honeycomb lattice. In comparison, 2D monolayers with the triangular lattice symmetry was much less explored. Recent studies nevertheless highlighted the potential of 2D triangular lattices to host new topological quantum states beyond 2D TIs. Zhang \etal~predicted the presence of ``destructive interference effect" between topological states in 2D triangular lattices with a tight-binding model~\cite{Zhang18p125430}. Feng \etal~reported the presence of 2D Dirac nodal-line fermions in monolayer Cu$_2$Si where Cu and Si atoms are arranged in a triangular lattice~\cite{Feng17p1007}. Another example is the non-centrosymmetric bulk compound PbTaSe$_2$ in which the triangular-lattice Pb monolayer plays an important role for the emergence of Wyel nodal lines in the presence of SOC~\cite{Bian16p10556}.  Notably, the 2D superconductivity was observed in a single layer of Pb grown on a Si(111) substrate, and the Pb atoms form a distorted triangular lattice~\cite{Zhang10p104,Noffsinger11p421}. 
    
In this work, we focus on 2D triangular lattices consisted of group-IV elements. Using first-principles density functional theory (DFT) calculations, we start by examining the electronic structures of a generic model: a single layer of Pb atoms arranged in a triangular lattice. It is found that the flat Pb monolayer supports a spinless nodal line in the absence of SOC while it becomes a trivial semimetal when the effect of SOC is taken into account. Interestingly, buckled Pb monolayer can also host a glide-mirror-protected nodal line in the absence of SOC. With SOC, the system becomes fully gapped and the gap size depends strongly on the degree of buckling. We further design a new compound PbHfO$_2$ by alternately stacking Pb and 1T-HfO$_2$ 2D sheets. The obtained quasi-3D crystal is dynamically stable and largely retains the electronic properties of Pb monolayers. By applying epitaxial strains to PbHfO$_2$, it is possible to drive an insulator-to-metal transition coupled with an anti-ferroelectric-to-paraelectric phase transition. The electronic structures of isostructural analogues, GeHfO$_2$ and SnHfO$_2$, are also studied. We conclude the the 2D triangular lattice may serve as a complimentary platform to design new Xenes with diverse properties and functions. 
\raggedbottom
\section{Computational Methods}
First-principles calculations are carried out with density functional theory (DFT) using the projector augmented wave (PAW) method implemented in the Vienna Ab initio Simulation (\texttt{VASP}) package~\cite{Blochi94p17953,Kresse96p11169}. The energy cutoff of the plane wave basis set is 760 eV, and a 10$\times$16$\times$6 Gamma-centered $k$-point mesh is used for Brillouin zone sampling. The lattice constants and atomic positions are fully optimized until the residual force on each atom is less than 0.001 eV/\AA. We carefully compared the optimized structures of PbHfO$_2$ using different exchange-correlation energy functionals: local density approximation (LDA)~\cite{Perdew81p5048}, generalized gradient approximation (GGA) with the Perdew-Burke-Ernzerhof (PBE) parameterization~\cite{Perdew96p3865}, PBE with Grimme dispersion corrections (PBE-D3)~\cite{Grimme10p154104}, optB88-vdW, and optB86b-vdW~\cite{Dion04p24,RomanPerez09p9,Thonhauser07p12,Klimes09p022201,Klimes11p19}. We find that LDA, PBE-D3, optB88-vdW, and optB86b-vdW predict similar structural parameters whereas PBE strongly overestimates the lattice constants. Therefore, all first-principles calculations reported here are performed using LDA. To reveal the spinless nodal line in the momentum space, the energy gap between selected pair of bands is calculated within the tight-binding (TB) scheme using \texttt{WannierTools}~\cite{Wu17pS0010465517303442}. The maximum localized Wannier functions (MLWF) TB Hamiltonian with Pb-$5p$ orbitals as projectors is constructed with \texttt{Wannier90} interfaced with \texttt{VASP}~\cite{Marzari97p12847}. The $Z_2$ invariant for the flat Pb monolayer including the effects of SOC is computed by tracking the evolution of hybrid Wannier functions using \texttt{Z2Pack}~\cite{Gresch17p075146}.

\section{Results and Discussions}
\subsection{Electronic structure of flat Pb triangular lattice}
\begin{figure}[ht]
\centering
\includegraphics[scale=1.0]{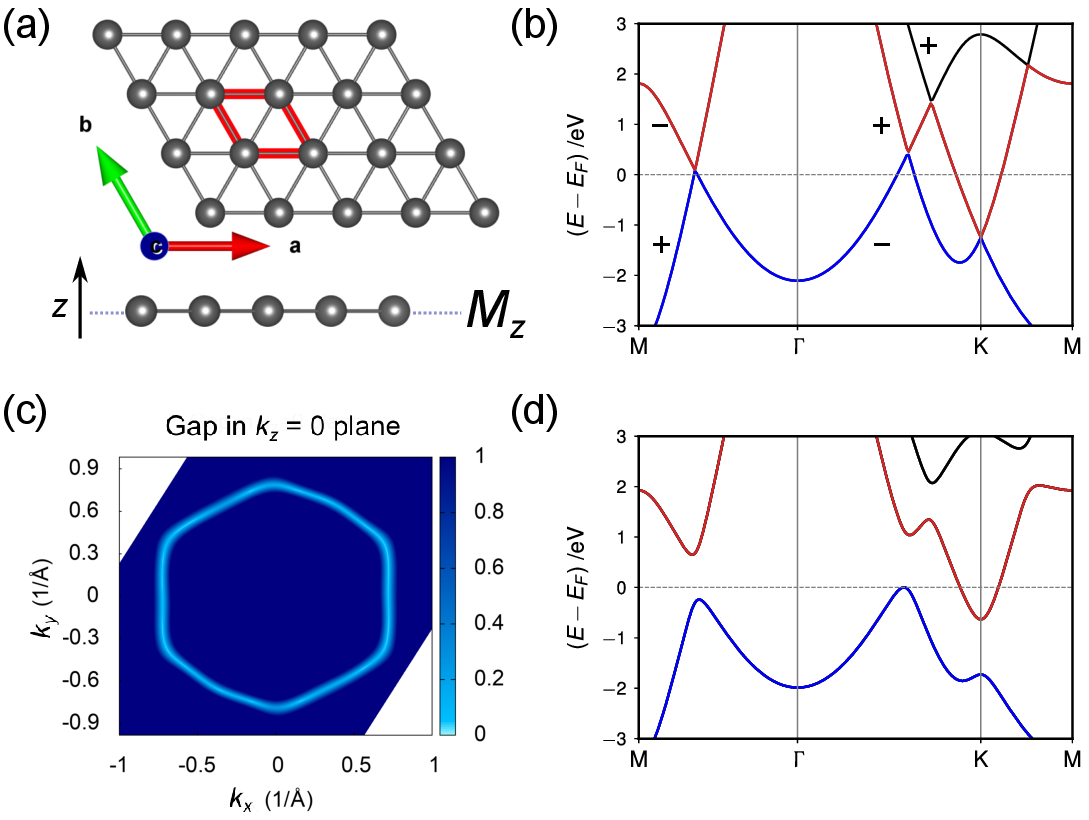}
 \caption{Electronic structures of flat Pb triangular lattice. (a) Crystal structure of Pb triangular lattice with mirror symmetry~\Mz. The 2D hexagonal unit cell is highlighted in red. (b) Electronic band structure without SOC along high-symmetry lines of the Brillouin zone of the hexagonal unit cell. (c) Spinless nodal loop in the $k_z=0$ plane. (d) Electronic band structure with SOC. The topmost valence and lowest conduction band are colored in blue and red, respectively.}
  \label{flatPb}
 \end{figure}

The free-standing triangular-lattice Pb monolayer has a hexgonal unit cell with an in-plane lattice constant of 3.18~\AA~(optimized without SOC, Fig.~\ref{flatPb}a). As all Pb atoms are coplanar, the mirror reflection symmetry with respect to the $xy$ plane (\Mz) is a natural consequence. The computed band structure without SOC is presented in Fig.~\ref{flatPb}b. There are two crossing points emerging along the high-symmetry path of M$-\Gamma-$K near the Fermi level ($E_F$), implying a possible closed loop on the Fermi surface. Figure~\ref{flatPb}c shows the momentum distribution of gapless nodal points in the $k_z=0$ plane, confirming a spinless Dirac nodal loop in the absence of SOC. To check whether the Dirac nodal loop is symmetry protected, we calculate the eigenvalues of \Mz~for the two bands near $E_F$.  It is found that the two bands have opposite \Mz~parities as marked by 
$+1$ and $-1$ in Fig.~\ref{flatPb}b. The opposing \Mz~parities indicate that the two bands belong to different irreducible representations (IRs) and will not hybridize to open a gap. 

The flat Pb monolayer has both time reversal symmetry ($\mathcal{T}$) and inversion symmetry ($\mathcal{P}$). In the presence of SOC and combined symmetry $\mathcal{PT}$, each band is doubly degenerate. Moreover, the Kramers pair will have opposing $\mathcal{M}_z$ parities. This can be seen from $\mathcal{M}_z\mathcal{PT}\psi_{+} =\mathcal{PT} \mathcal{M}_z\psi_{+} =\mathcal{PT}i\psi_{+} = -i\mathcal{PT}\psi_{+}$, which implies the Kramers pair $\psi_{+}$ and $\mathcal{PT}\psi_{+}$ have opposite mirror eigenvalues ($+i$ and $-i$). As a result, the two doublet Bloch states forming the nodal loop in the mirror invariant plane ($k_z=0$) will hybridize and anticross. As shown in the band structure computed with SOC (Fig.~\ref{flatPb}d), the bands along M$-\Gamma-$K$-$M are fully gapped.  Because of the large SOC strength afforded by Pb atoms, the size of SOC gap is quite large ($\approx$1.1~eV). Similar to previous study of free-standing Cu$_2$Si where mirror-protected Dirac nodal lines are gapped after including SOC~\cite{Feng17p1007}, here we also see the annihilation of Dirac nodal line in the presence of SOC. We calculate the $Z_2$ topological number and find that the flat Pb monolayer is a trivial semimetal (0:000) with an electron pocket around the high-symmetry point K (Fig.~\ref{flatPb}d).

\subsection{Electronic structure of buckled Pb triangular lattice}
\begin{figure}[ht]
\centering
\includegraphics[scale=1.0]{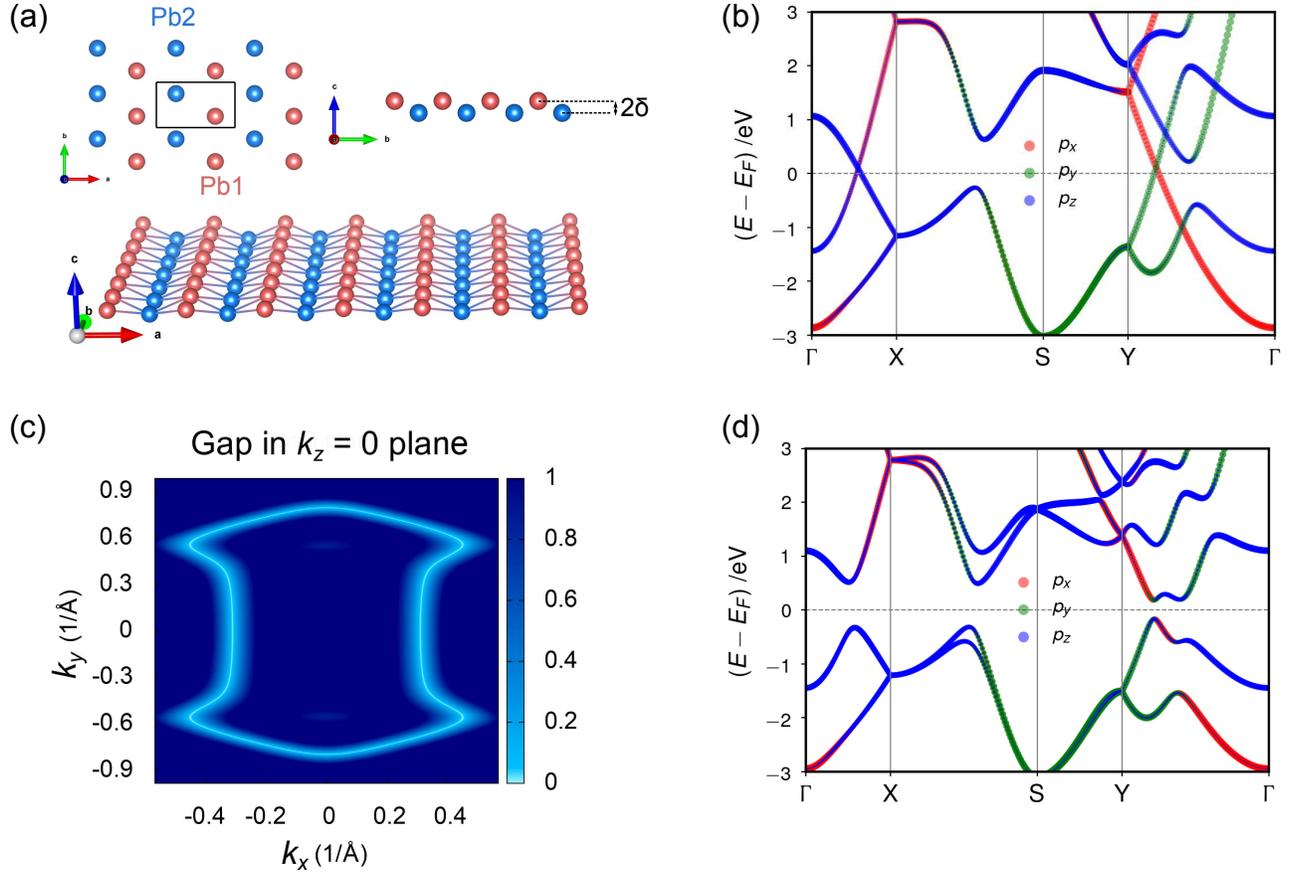}
\caption{Electronic structures of buckled Pb triangular lattice. (a) Crystal structure of buckled Pb sheet. The orthorhombic unit cell has two atoms colored in red and blue respectively. (b) Orbital-resolved electronic band structure without SOC along high-symmetry lines of the Brillouin zone of the orthorhombic unit cell. (c) Spinless nodal loop in the $k_z=0$ plane. (d) Orbital-resolved band structure with SOC. }
  \label{bPb}
 \end{figure}

It is well known that the atomic buckling can strongly affect the physical and chemical properties of 2D materials~\cite{Molle17p163}. We investigate the electronic structures of buckled Pb monolayer in which one column of Pb atoms (Pb1, colored in red) are shifted upwards while adjacent columns of Pb atoms (Pb2, colored in blue) are shifted downwards (Fig.~\ref{bPb}a). This buckling pattern resembles 2D boron sheets grown on silver surfaces~\cite{Mannix15p1513}.
The primitive unit cell of the buckled sheet contains two Pb atoms and has an orthorhombic lattice in the space group of $Pmmn$ ($a=$5.41~\AA~and $b=3.18$~\AA, optimized without SOC). The parameter $\delta$ that measures the degree of bucking is defined as $\delta=\frac{1}{2}(z^{\rm Pb1} - z^{\rm Pb2})$ where $z^{\rm Pb}$ is the $z$-component of Pb position. One might expect the spinless nodal loop observed in the flat configuration would become gapped after the removal of the mirror reflection symmetry \Mz~due to buckling. Interestingly, we identify a gapless nodal loop in buckled Pb monolayer ($\delta=0.35$~\AA) as well. The orbital-resolved band structure reveals Dirac-cone-like linear crossing points along high-symmetry lines of $\Gamma-$X and $\Gamma-$Y near $E_F$ (Fig.~\ref{bPb}b), corresponding to an anisotropic Dirac nodal loop in the $k_z=0$ plane (Fig.~\ref{bPb}c). This implies the existence of some symmetry other than \Mz~ensuring gapless crossing at any arbitrary $k$-points lying in the plane of $k_z=0$. It turns out the space group $Pmmn$ has a glide mirror plane $\{\mathcal{G}_{z}|\frac{1}{2} \frac{1}{2}\}$, which can also host nodal lines in glide-invariant planes in the momentum space.

Similar to the flat monolayer, the nodal line becomes fully gapped in the buckled Pb sheet after including SOC. This can be understood by following the analysis in the work of Fang {\em et al}~\cite{Fang15p081201}. For a spinfull system, the bands in the glide-invariant plane $k_z=0$ can be labeled by their \Gz~eigenvalues $g_{\pm}={\pm}ie^{(ik_x+ik_y)/2}$. In the presence of $\mathcal{PT}$, all bands are two-fold degenerate at every $k$-point. Suppose at $(k_x, k_y, 0)$, $\psi_{+}$ satisfies $\mathcal{G}\psi_{+}=g_+\psi_+$, then $\mathcal{G}_z\mathcal{PT}\psi_{+} = e^{i(k_x+k_y)}\mathcal{PT}\mathcal{G}_z\psi_{+} = e^{i(k_x+k_y)}\mathcal{PT} ie^{i(k_x+k_y)/2}\psi_{+} = g_{-}\mathcal{PT}\psi_{+}$, which implies the doubly-degenerate bands related by $\mathcal{PT}$ form a Kramers pair while having opposite $\mathcal{G}_z$ eigenvalues in the $k_z=0$ plane. Therefore, the bands forming the nodal line are allowed to hybridize and open a gap when the SOC is considered (Fig.~\ref{bPb}d). We find that the bucked Pb monolayer is a semiconductor with a band gap of 0.3~eV. 
Noted that all bands are four-fold degenerate at X, Y, and S, which are Dirac points protected by two screw axes $\{\mathcal{C}_x|\frac{1}{2}0\}$ and $\{\mathcal{C}_y|0\frac{1}{2}\}$~\cite{Young15p126803}. In principle, one may tune the Fermi level (via hole or electron doping) to expose these Dirac points, making the system a 2D Dirac semimetal. 

\begin{figure}[ht]
\centering
\includegraphics[scale=1.0]{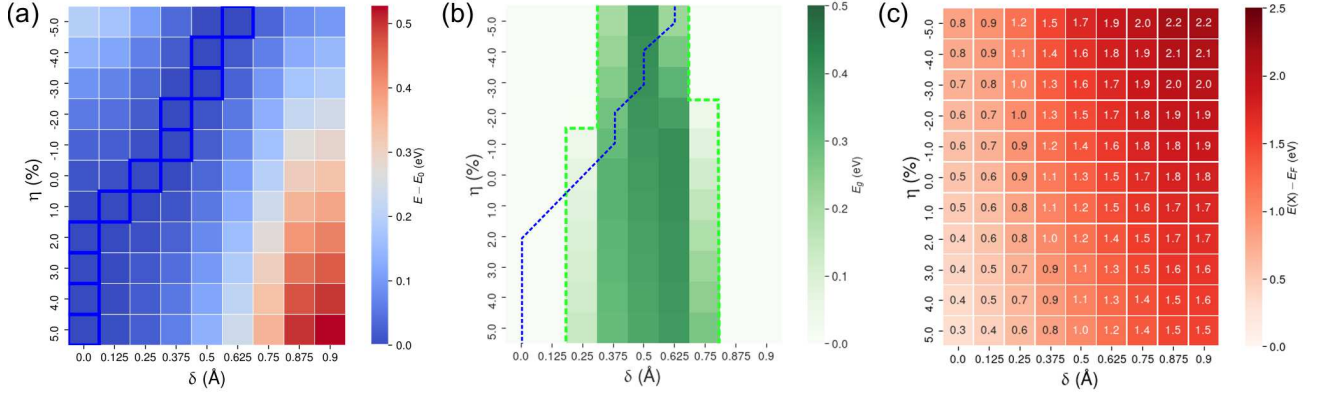}
\caption{ Structural and electronic properties of Pb triangular lattice as a function of in-plane strains ($\eta$) and buckling parameters ($\delta$). 
(a) Relative energies as a function of $\eta$ and $\delta$. At each in-plain strain state, the configuration with the lowest energy ($E_0$, highlighted by a blue square) is chosen as the reference for the calculations of energy differences ($E-E_0$).
(b) Band gaps ($E_g$) as a function of $\eta$ and $\delta$. The metal-insulator transition boundaries are highlighted by dashed lines in green.
The dashed blue line tracks the lowest-energy configurations in (a), and reveals an insulator-to-metal transition with increasing tensile strains.
(c) Energy difference between the four-fold degenerate Dirac point at the high-symmetry $k$-point X and the Fermi level ($E_F$) with respect to $\eta$ and $\delta$.}
  \label{heat}
 \end{figure}

The finding that the flat Pb triangular lattice is a semimetal whereas its buckled sheet is a semiconductor suggests a metal-to-insulator transition driven by the atomic buckling. We calculate the energies and band gaps of Pb monolayer for different combinations of in-plane strains ($\eta$) and buckling parameters using an orthorhombic unit cell ($a=\sqrt{3}b$). The in-plain strain is defined as $\eta=(b-b_0)/b_0$, where $b_0$ is the equilibrium Pb-Pb distance in the flat monolayer. As expected, a compressive strain will favor a buckled configuration while a tensile strain will stabilize the flat configuration (Fig.~\ref{heat}a). From the calculated band gaps shown in Fig.~\ref{heat}b, it is clear that the band gap depends sensitively on the degree of buckling, and an insulator-to-metal transition can be realized by applying tensile stains. These results indicate the Pb monolayer allows for great tunability of the electronic properties via strain engineering.  Moreover, we compute the energy difference between the Dirac point at X point (Fig.~\ref{bPb}d) and the Fermi level (Fig.~\ref{heat}c) and find that the Dirac point is approaching the Fermi surface with increasing in-plain tensile strains. 

\subsection{Electronic structures of Ge and Sn triangular lattices}
\begin{figure}[ht]
\centering
\includegraphics[scale=1.3]{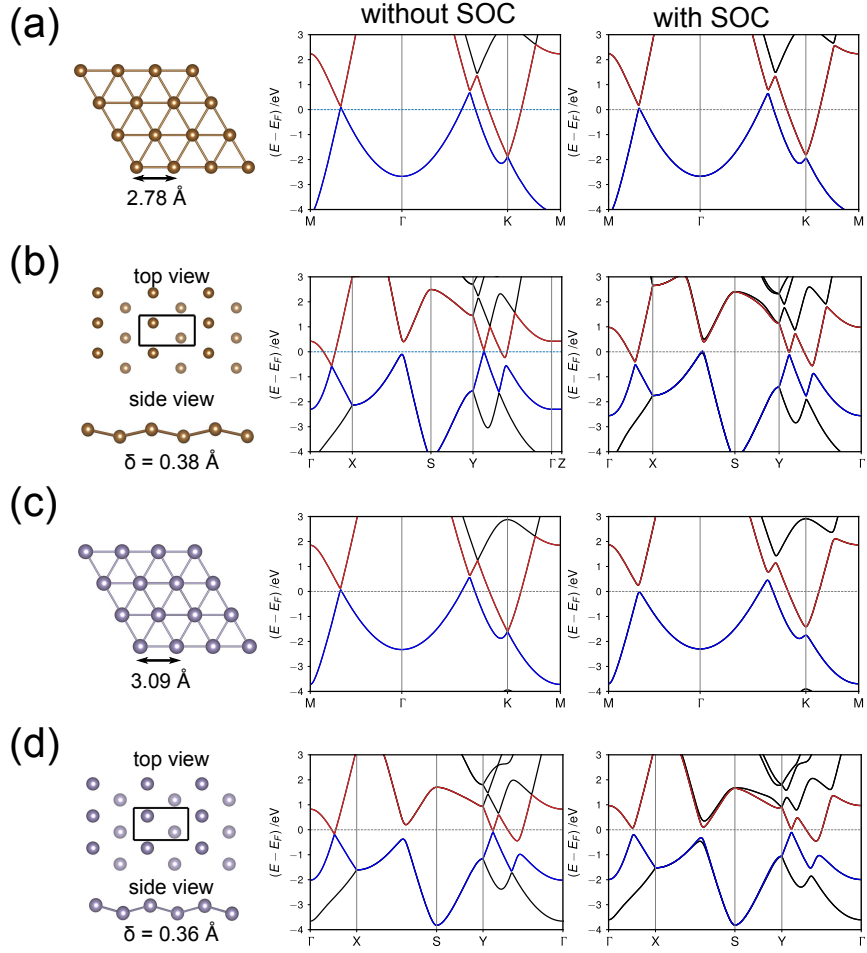}
 \caption{Electronic band structures for (a) flat and (b) buckled Ge monolayer and (c) flat and (d) buckled Sn monolayer in a triangular lattice.}
  \label{GeSnML}
 \end{figure}

Figure~\ref{GeSnML} reports the optimized structures and electronic band structures for flat and buckled Ge and Sn triangular lattices. The band dispersions of flat Ge and Sn monolayers obtained without SOC closely resemble those in the flat Pb sheet, both indicating the presence of spinless nodal loops in the $k_z=0$ plane.  Similarly, buckled Ge and Sn monolayers also host nodal lines in glide-invariant planes in the momentum space, respectively. After considering SOC, the nodal lines become gapped. Because of the weak SOC strengths of Ge and Sn atoms, their buckled sheets show the features of semimetals, different from the semiconducting behavior of a buckled Pb triangular lattice. 
As the gaps at the anti-crossing points near $E_F$ are relatively small and band dispersions remain largely linear, these two systems may act as 2D quasi-free-fermions in practice~\cite{Jin17pTong, Feng17p1007}.

\subsection{Design 3D materials supporting Pb monolayer}
\begin{figure}[ht]
\centering
\includegraphics[scale=1.3]{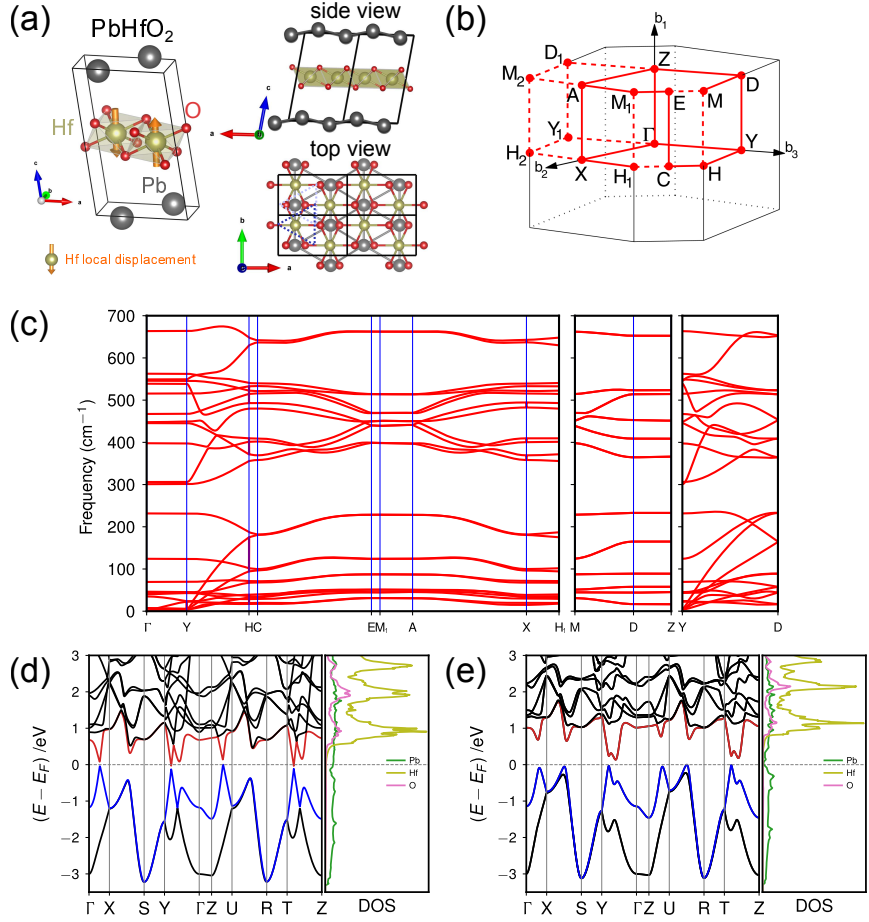}
 \caption{Electronic structures of PbHfO$_2$. (a) Crystal structure of PbHfO$_2$ in the space group of $P2_1/m$. (b) Brillouin zone of a monoclinic unit cell. (c) Phonon spectrum along high-symmetry lines of the Brillouin zone of the monoclinic unit cell. Electronic band structures and projected density of states (DOS) (d) without SOC and (e) with SOC. The topmost valence and lowest conduction band are colored in blue and red, respectively.}
  \label{PHO}
 \end{figure}

So far, we have treated group-IV triangular lattices as generic model systems. It is natural to ask whether it is feasible to access their tunable electronic structures in realistic materials. Our phonon calculations of the Pb monolayer reveal vibrational modes with imaginary frequencies, indicating it is dynamically unstable.  We propose to realize a quasi-3D material by stacking Pb triangular lattice with other ``electrically inert" 2D materials in which the interlayer coupling is weak. With this design principle, we identify 1T-HfO$_2$ monolayer~\cite{Weng18p26453} as a suitable building block due to its wide band gap (6.73~eV) and lattice symmetry compatible with a triangular lattice. Similar to 2D transition metal dichalcogenides, the 1T-HfO$_2$ monolayer has two layers of hexagonal oxygen lattices sandwiching a slab of hexagonally packed Hf lattice, and each atomic layer has a triangular lattice symmetry. By alternately stacking Pb and 1T-HfO$_2$ monolayers, we obtain a quasi-3D material PbHfO$_2$. Figure~\ref{PHO}a shows the optimized unit cell of PbHfO$_2$ which has a monoclinic space group $P2_1/m$ and lattice constants of $a=5.492$~\AA, $b=3.187$~\AA, $c=9.118$~\AA, $\beta=96.7^\circ$, $\alpha = \gamma=90^\circ$. The bucking parameter $\delta$ of the Pb layer in PbHfO$_2$ is 0.6~\AA. Interestingly, PbHfO$_2$ is antiferroelectric as Hf atoms are displaced with respect to the center of their surrounding oxygen octahedra while the two Hf atoms in the unit cell have opposite local displacements (Fig.~\ref{PHO}a). The structure of PbHfO$_2$ is confirmed stable through computations of the phonon dispersions along high-symmetry lines of the Brillouin zone of the monoclinic unit cell (Fig.~\ref{PHO}b). As shown in Fig.~\ref{PHO}c, the phonon spectrum has no imaginary frequencies, demonstrating the designed material PbHfO$_2$ is indeed dynamically stable. Additionally, the dispersions along high-symmetry lines along $k_z$ such as $\Gamma-$Y, A$-$M$_1$, X$-$H$_1$, and Z$-$D are nearly flat, indicating weak interactions between Pb and HfO$_2$ layers. 

The electronic band structure and projected density of states (DOS) of PbHfO$_2$ in the absence of SOC are depicted in Fig.~\ref{PHO}d. To  compare directly with the band structure of a freestanding Pb monolayer. we plot the band dispersions along high-symmetry lines of the Brillouin zone of the primitive orthorhombic lattice.
Clearly, the band dispersions near $E_F$ bare strong similarity to those in the buckled Pb sheet (Fig.~\ref{bPb}b). We find that the states close to the Fermi level are mainly from Pb-$5p$ orbitals with negligible contributions from Hf and O atoms, supporting our design principle that the HfO$_2$ layer is ``electrically inert". The most notable change in the band structure of PbHfO$_2$ is the anti-crossing along $\Gamma-$X and $\Gamma-$Y (Fig.~\ref{PHO}d). This is expected given that the $P2_1/m$ space group no longer hosts the mirror (\Mz) or the glide mirror (\Gz) symmetry. After considering SOC, PbHfO$_2$ is a semiconductor with a band gap of 0.14~eV (Fig.~\ref{PHO}e). We add that the $P2_1/m$ space group still has a screw axis $\{\mathcal{C}_y|0\frac{1}{2}\}$, which protects the four-fold degenerate Dirac points at Y and S. However, these Dirac points are quite far away from the Fermi level, making it difficult to access their topological properties. 

\subsection{Strain engineering of PbHfO$_2$}
Advances in thin-film epitaxy taking advantage of misfit strains between the film and the underlying substate have been widely used to finely control the structure, response, and properties of functional materials such as ferroelectrics~\cite{Schlom07p589, Damodaran16p263001,Xu15p79}, topological quantum materials~\cite{Liu14p294,Liu16p1663}, and low-dimenstional materials~\cite{Xu20p3141, Li20p1151,Zhao20p046801}. Our previous analysis of Pb monolayer has already shown that the band gaps depend strongly on the degree of buckling, which can be tuned by applying epitaxial strains. 

\begin{figure}[ht]
\centering
\includegraphics[scale=1.0]{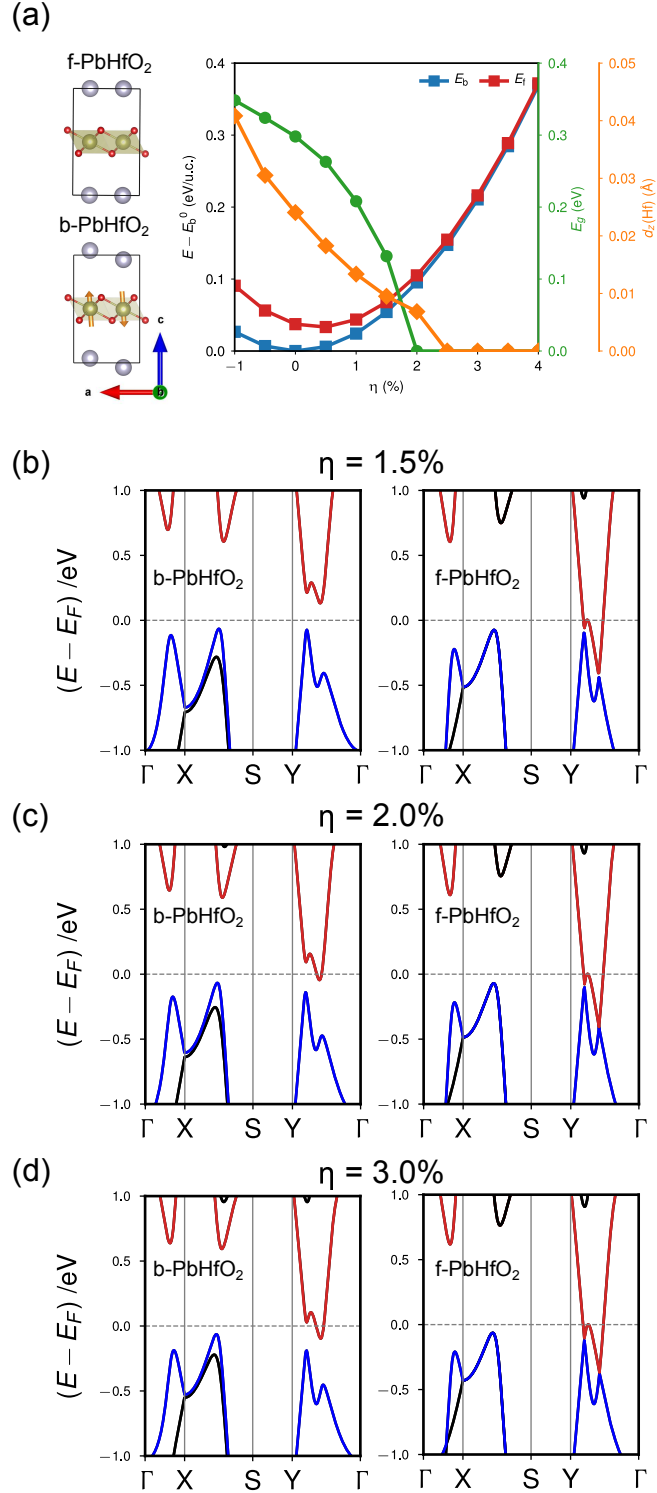}
 \caption{ (a) Evolution of energies, band gaps ($E_g$), and Hf local displacements ($d_z({\rm Hf})$) as a function of in-plane strains $\eta$. The energy of the ground state of PbHfO$_2$ ($E_b^0$) is chosen as the zero energy reference. $E_{\rm b}$ and $E_{\rm f}$ are the energies of b-PbHfO$_2$ and f-PbHfO$_2$, respectively. Band structures of b-PbHfO$_2$ (left) and f-PbHfO$_2$ (right) at (b) $\eta=1.5\%$, (c) $\eta=2.0\%$, and (d) $\eta=3.0\%$}.
  \label{PHO_strain}
 \end{figure}

We investigate the effects of in-plane deformations along the $a$ and $b$ axes by rescaling the lattice vectors of PbHfO$_2$ while allowing the atomic positions and the length of $c$ axis to fully relax (the optimized structure is denoted as b-PbHfO$_2$). At each in-plain strain condition, we also carry out a constrained optimization of PbHfO$_2$ by forcing Pb atoms coplanar, denoted as f-PbHfO$_2$ (see illustration in Fig.~\ref{PHO_strain}a). This allows the determination of the stability of b-PbHfO$_2$ with a buckled Pb layer relative to f-PbHfO$_2$ as well as the critical strain at which a buckled-to-flat transition may occur.  Figure~\ref{PHO_strain}a reports the evolution of the energies of b-PbHfO$_2$ and f-PbHfO$_2$ ($E_{\rm b}$ vs $E_{\rm f}$), band gap ($E_g$), and local displacement of Hf ($d_z$(Hf)) as a function of in-plain strains. It is found that at a critical tensile strain of $\approx2$~\%, the energies of b-PbHfO$_2$ and f-PbHfO$_2$ are nearly identical, and the system becomes metallic in both configurations. This strain-driven insulator-metal transition is coupled with an antiferroelectric-paraelectric phase transition as $d_z$(Hf) decreases with increasing tensile strains. We show the SOC band structures of b-PbHfO$_2$  and f-PbHfO$_2$ at some selective strain states in Fig.~\ref{PHO_strain}b-d. 
\vspace{-0.5cm}
\subsection{Electronic structure of SnHfO$_2$}
\begin{figure}[ht]
\centering
\includegraphics[scale=1.5]{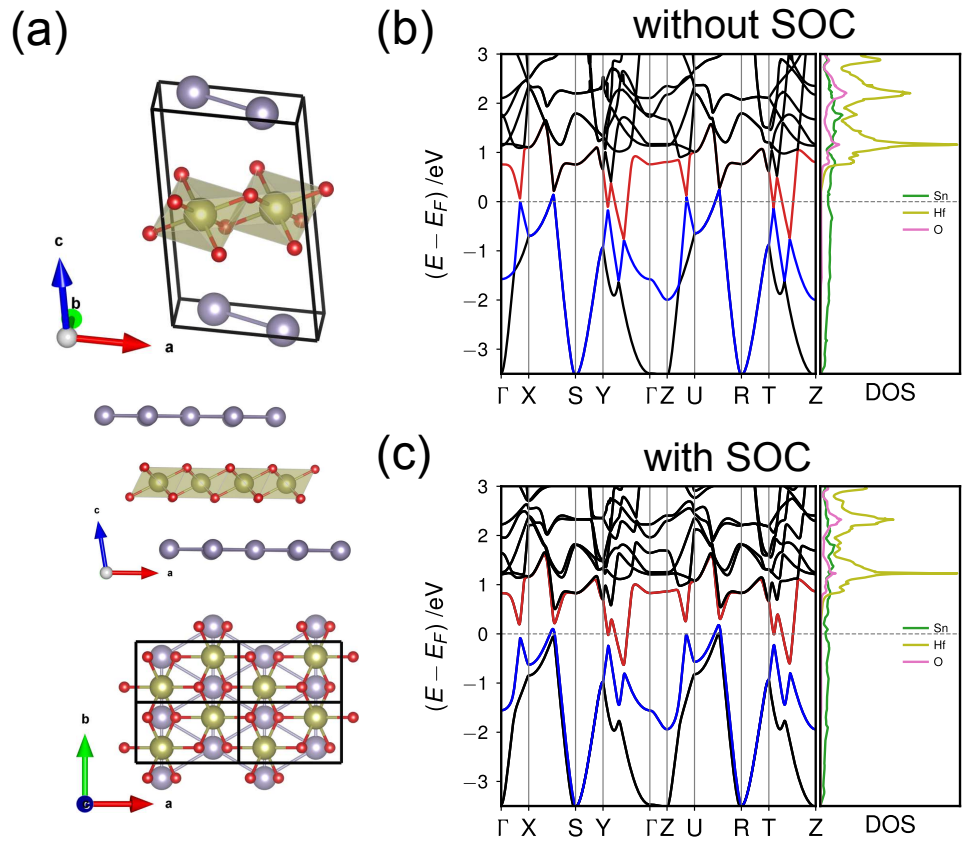}
 \caption{Electronic structures of SnHfO$_2$. (a) Optimized structure of SnHfO$_2$ in the space group of $P2_1/m$. Electronic band structure (b) without SOC and (c) with SOC}
  \label{SHO}
 \end{figure}

The isostructural analogues of PbHfO$_2$ are studied by replacing Pb with Ge and Sn, respectively. Due to the large lattice mismatch between HfO$_2$ and Ge monolayers, Ge atoms in the optimized GeHfO$_2$ form isolated 1D-channels. In comparison, the Sn atoms in SnHfO$_2$ are nearly coplanar and are arranged in a triangular lattice (Fig.~\ref{SHO}a). The band structure in the scalar-relativistic case reveals multiple points resembling Dirac crossings along Y$-\Gamma-$X$-$S and  Z$-$U$-$R$-$T~Fig.~\ref{SHO}b). However, because SnHfO$_2$ is in the space group of $P2_1/m$ which does not contain \Mz~or \Gz, close examinations show that the topmost valence and the lowest conduction bands of SnHfO$_2$ are already gaped. The flat bands along $\Gamma-$Z again confirm weak interactions between Sn and HfO$_2$ layers. When the SOC is taken into account, the electronic spectrum of SnHfO$_2$ further gaps out (Fig.~\ref{SHO}c). The four-fold degeneracy at Y and S points remain protected due to the screw axis $\{\mathcal{C}_y|0\frac{1}{2}\}$.

 \section{Conclusions}
First-principle density functional theory calculations are employed to explore the structural and electronic properties of 2D triangular lattices made of group-IV elements. Unlike Xenes with a honeycomb symmetry,  2D monoelemental crystal with a triangular lattice is mostly overlooked by the mainstream research. This work aims to fill in the gap and to encourage further experimental and theoretical studies on this new type of broadly-defined Xene. Our results demonstrate that these 2D triangular lattices may host rich physics, ranging from nonsymmorphic-symmetry-protected nodal lines and Dirac points to strain-driven metal-insulator transitions. The designed quasi-3D crystal PbHfO$_2$ is predicted to be dynamically stable and largely retains the electronic properties of 2D Pb triangular lattice. The emergence of anti-ferroelectricity in PbHfO$_2$ opens up opportunities for electric field-control of the buckling of Pb monolayer and its band gaps. The design principle proposed here to alternatively stacking electrically inert HfO$_2$ monolayer and 2D Pb/Sn monolayers can also be applied to known Xenes such as stanene and plumbene, offering a useful platform to design and manipulate emergent phenomena in Xenes.

 \section{Acknowledgments}
XD and SL acknowledge the support from Westlake Foundation. The computational resource is provided by Westlake Supercomputer Center. 

\bibliography{SL}
   \newpage

   \newpage
                           
\newpage
 
 \newpage                                                

\newpage
 
\newpage

\newpage
\end{document}